\documentclass[sigconf]{acmart}

\renewcommand\footnotetextcopyrightpermission[1]{} 
\usepackage{caption}
\usepackage{subcaption}

\AtBeginDocument{%
  \providecommand\BibTeX{{%
    \normalfont B\kern-0.5em{\scshape i\kern-0.25em b}\kern-0.8em\TeX}}}

\setcopyright{acmcopyright}
\copyrightyear{2020}
\acmYear{2018}
\acmDOI{10.1145/1122445.1122456}

\acmConference[Conference'20]{}{June 2020}

\begin{document}


\title{Learning Effective Representations for Person-Job Fit by Feature Fusion
}

\author{Junshu	Jiang}
\email{jiangjunshu516@pingan.com.cn}
\affiliation{%
  \institution{Ping An HR-X Laboratory}
    \city{Shen Zhen}
  \country{China}
}

\author{Songyun	Ye}
\email{yesongyun678@pingan.com.cn}
\affiliation{%
  \institution{Ping An HR-X Laboratory}
  \city{Shen Zhen}
  \country{China}
}

\author{Wei Wang}
\email{wangwei@comp.nus.edu.sg}
\affiliation{%
  \institution{National University of Singapore}
  \country{Singapore}}

\author{Jingran	Xu}
\email{xujingran131@pingan.com.cn}
\affiliation{%
  \institution{Ping An HR-X Laboratory}
    \city{Shen Zhen}
  \country{China}
}

\author{Xiaosheng	Luo}
\authornote{Corresponding Author.}
\email{luoxiaosheng944@pingan.com.cn}
\affiliation{%
  \institution{Ping An HR-X Laboratory}
    \city{Shen Zhen}
  \country{China}
}

\renewcommand{\shortauthors}{Jiang, et al.}


\begin{abstract}
Person-job fit is to match candidates and job posts on online recruitment platforms using machine learning algorithms. 
The effectiveness of matching algorithms heavily depends on the learned representations for the candidates and job posts. 
In this paper, we propose to learn comprehensive and effective representations of the candidates and job posts via feature fusion. First, in addition to applying deep learning models for processing the free text in resumes and job posts, which is adopted by existing methods, we extract semantic entities from the whole resume (and job post) and then learn features for them. By fusing the features from the free text and the entities, we get a comprehensive representation for the information explicitly stated in the resume and job post. Second, however, some information of a candidate or a job may not be explicitly captured in the resume or job post. Nonetheless, the historical applications including accepted and rejected cases can reveal some implicit intentions of the candidates or recruiters. Therefore, we propose to learn the representations of implicit intentions by processing the historical applications using LSTM. Last, by fusing the representations for the explicit and implicit intentions, we get a more comprehensive and effective representation for person-job fit.
Experiments over 10 months real data show that our solution outperforms existing methods with a large margin. Ablation studies confirm the contribution of each component of the fused representation. The extracted semantic entities help interpret the matching results during the case study.
\end{abstract}



\maketitle

\section{Introduction}
Online recruitment platforms, e.g., LinkedIn,
make it easy for companies to post jobs and for job seekers to submit resumes. In recent years, the number of both job posts and resumes submitted to online recruitment platforms is growing rapidly. For example, in U.S, there are over 3 million jobs posted on LinkedIn in every month\cite{linkedinreport}.
Traditionally, resumes submitted for each job are reviewed manually by the recruiter to decide whether to offer the candidates the job interview. However, manual reviewing is slow and expensive to handle the overwhelming new job posts and resumes on online platforms. It is essential to design effective algorithms to do job-resume matching automatically. This problem is called person-job fit. 

Multiple approaches have been proposed for person-job fit. Earlier solutions consider person-job fit as a recommendation problem and apply collaborative filtering (CF) algorithms~\cite{iscid2014,DBLP:conf/asunam/DiabyVL13,DBLP:conf/www/LuHG13}. However, CF algorithms ignore the content of the job post and the resume, e.g., the working experience of the candidate and the job requirement. In contrast, when we do manual reviewing, we read the resume to understand the candidate (e.g., the skills and experience); we read the job post to understand the requirements; then we make a decision, i.e., whether the candidate should be offered an interview. We can see that the content of the resume and job post plays a key role in person-job fit. It is thus vital to extract effective representation of the content.

Recently, deep learning models have largely improved the performance of natural language processing tasks, including semantic matching~\cite{DBLP:conf/cikm/HuangHGDAH13,DBLP:conf/www/Mitra0C17} and question answering. Deep-learning-based methods~\cite{DBLP:conf/cikm/LeHSZ0019,DBLP:conf/kdd/YanLSZZ019,DBLP:conf/sigir/QinZXZJCX18,bian-etal-2019-domain} are consequently introduced for person-job fit, focusing on learning effective representations of the free text of the job post and resume. The learned representations are then compared to generate a matching score. However, they only process the text paragraphs including the working experience and job requirements, and fail to comprehend other (semi-) structured fields like the education, skills, etc. This is partly because 
deep learning models are typically applied for natural language sentences instead of (semi-) structured fields. As a result, valuable information from these fields are left unexploited.

\begin{figure}[h]
    \begin{subfigure}{0.23\textwidth}
    \centering
    \includegraphics[width=\textwidth]{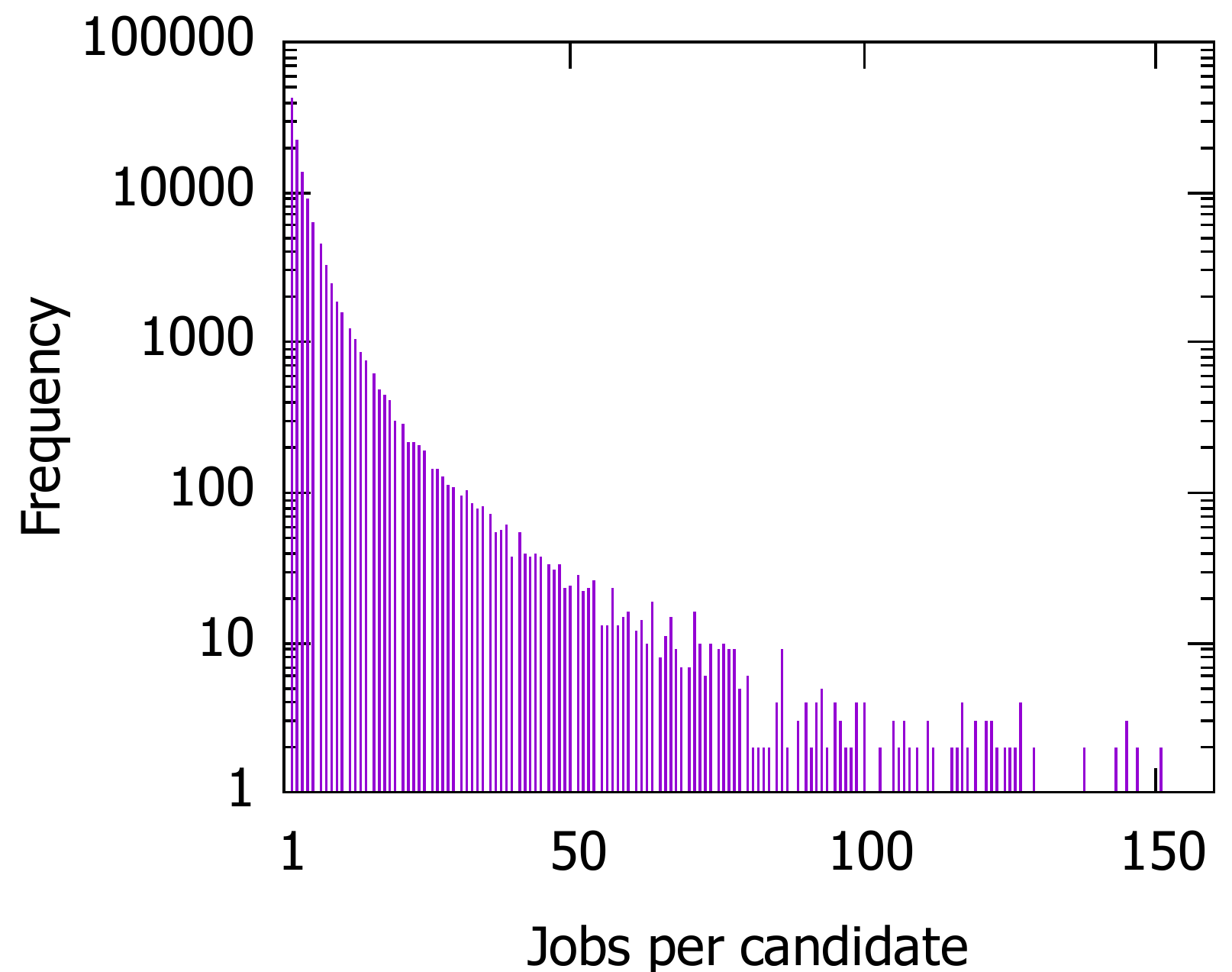}
    \caption{}
    \label{fig:jobspercand}
    \end{subfigure}
    \hfill
    \begin{subfigure}{0.23\textwidth}
    \centering
    \includegraphics[width=\textwidth]{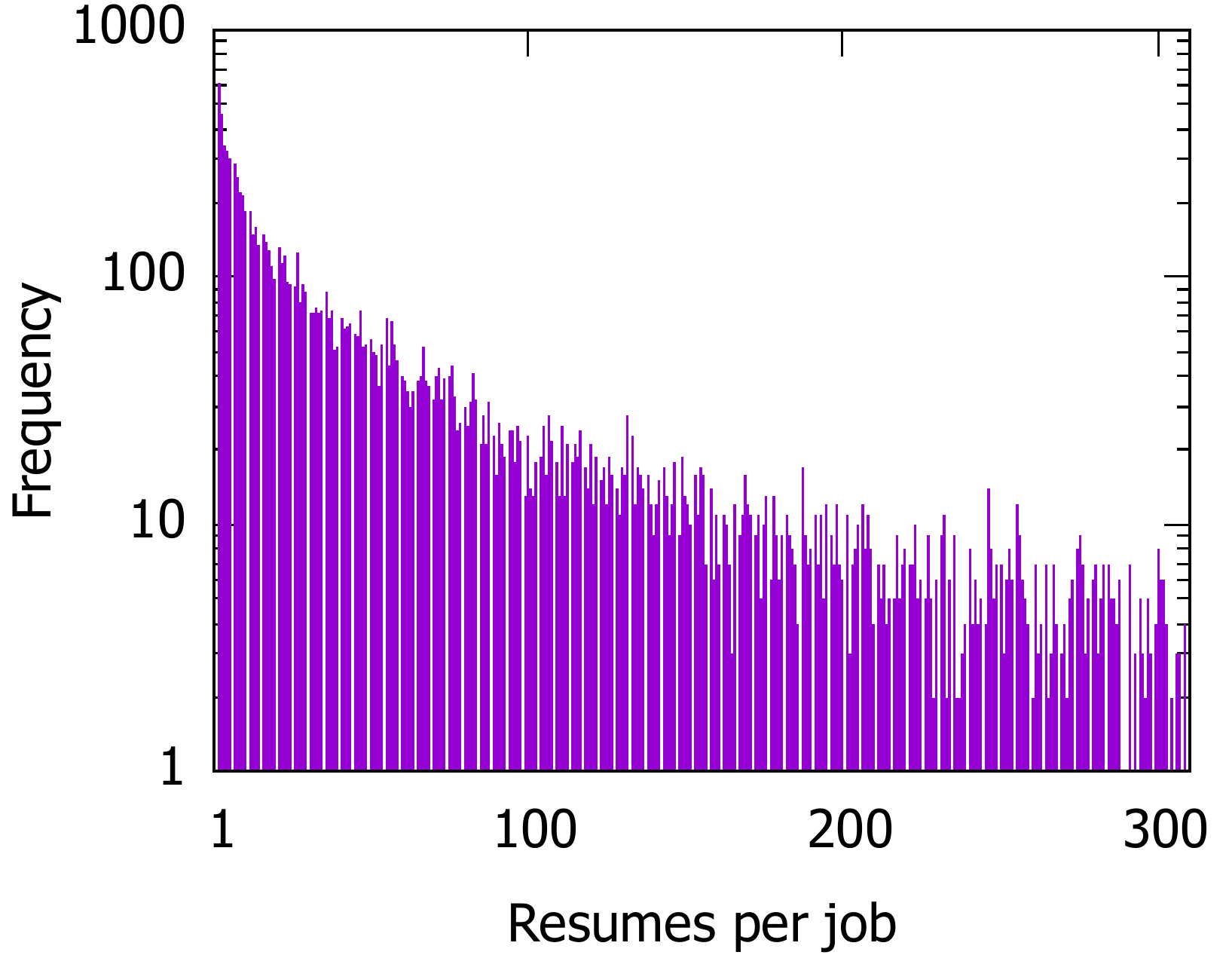}
    \caption{}
    \label{fig:resumesperjob}
    \end{subfigure}
    \caption{Frequency for the number of jobs applied per candidate (left) and resumes received per job (right).}
    \label{fig:history}
\end{figure}

\begin{figure*}[t!]
    \centering
    \includegraphics[width=0.98\textwidth]{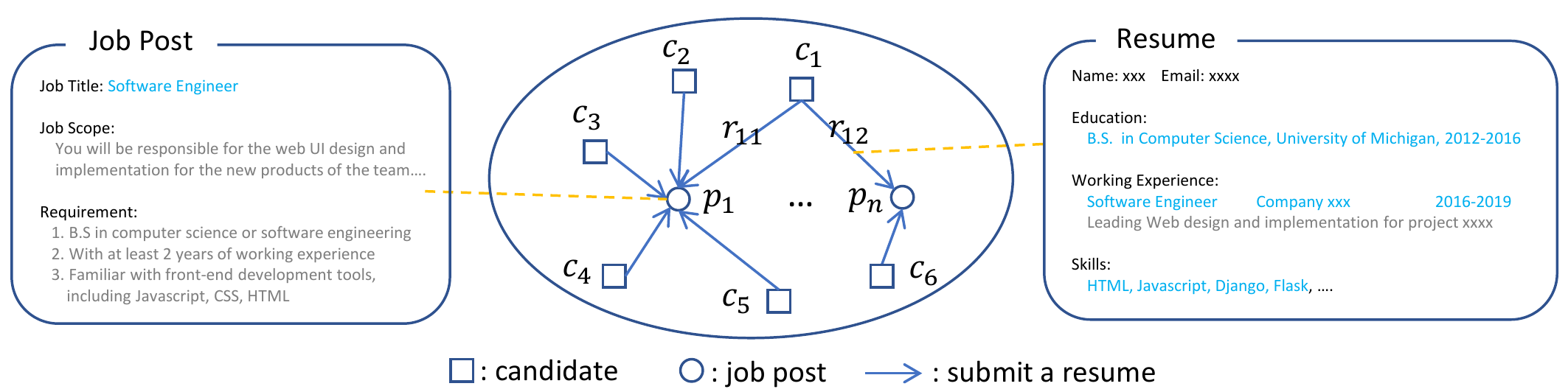}
    \caption{Illustration of candidates, job posts and resumes.}
    \label{fig:prob}
\end{figure*}

Moreover, most of existing deep-learning-based solutions ignore the historical applications of the candidate and the job post. It is common for a candidate to apply multiple jobs and for a job to receive multiple resumes, as shown in Figure~\ref{fig:history}. The numbers are derived from our experiment dataset. In specific, about 36\% have applied more than one jobs and about 88\% jobs have received more than one resumes. 
The history data could provide extra information of the candidate and job. Specifically, sometimes the job description is not written carefully or comprehensively, e.g., missing some requirements or unclear preference between two skills (Python versus C++); sometimes the recruiter's requirement or expectation may change dynamically, e.g., increasing if the received resumes so far are of very high quality. For such cases, the history data including accepted resumes and rejected resumes of a job could help to infer the recruiter's implicit intentions not elaborated in the job description.

In addition, deep learning models are typically difficult to interpret~\cite{DBLP:journals/corr/Lipton16a} due to complex internal transformations, although a lot of attention has been paid to this issue~\cite{8099837,DBLP:conf/iccv/SelvarajuCDVPB17,DBLP:conf/icml/KohL17}. As a result, deep-learning-based person-job fit solutions face the interpretation problem. In real deployment, yet, it is necessary to explain why a candidate is accepted or rejected for a given job post. 

Towards these issues, in this paper, we propose a feature fusion solution. First, we propose a semantic entity extraction step to extract semantic entities, including the university and working years, from the whole resume and job post. All extracted entities are then concatenated into a vector, which captures the high-level semantics of the content and is easy to interpret. The semantic vector is further transformed through an adapted DeepFM model to learn the correlations among the entities.
We also apply a convolutional neural network (CNN) over the text fields in the resume (resp. job post) following existing works. The outputs from DeepFM and CNN are fused via concatenation to produce a feature vector representing the explicit intention of a resume (resp. job post). 
Second, to exploit the history information, we propose a new encoding scheme for the job-resume pair from an application. All the historical applications, including both the accepted and rejected cases, of a candidate (resp. a job post) are processed by a LSTM model to learn the implicit intention. 
Last, we do a late fusion of the representations for the explicit and implicit intentions to represent the job and candidate comprehensively. 

Extensive experimental evaluations over real data show that our solution outperforms existing methods. We also conduct ablation studies to verify the effectiveness of each component of our solution. In addition, case studies are presented to demonstrate the contribution of semantic entities to model interpretation. Our solution has been deployed partially for one year. Some experience on improving the efficiency and reducing the cost will be shared at the end of this paper.

Our contributions include
\begin{enumerate}
    \item We propose a method to learn the representation for the explicit intention of the candidate and recruiter by processing the whole content of the resume and job post. 
    \item We propose a method to learn the representation for the implicit intention of the candidate and recruiter by aggregating the historical applications.
    \item We fuse the two sets of representations to get an effective and comprehensive representation for both the job post and candidate. Extensive experiments are conducted to confirm the effectiveness and interpretability of our proposed solution.
\end{enumerate}

\section{Related Works}

With the proliferation of online recruitment services, various recruitment analysis tasks have been studied, including career transition (a.k.a. talent flow) analysis~\cite{DBLP:conf/kdd/ChengXCACG13,DBLP:conf/www/WangZPB13,DBLP:journals/tkde/XuYYXZ19}, job (resume) recommendation and person-job fit. We shall review the recent papers on the last two tasks in more details.

\subsection{Job and Resume Recommendation}
Existing online recruitment platforms like LinkedIn typically have the user information including the demographic profile and working experience. They can also record the actions of the user on each job post, e.g., clicking and browsing time. Existing recommendation algorithms such as collaborative filtering have been adapted~\cite{1579569, DBLP:conf/www/LuHG13, 10.1145/2492517.2500266} to recommend jobs to users or recommend resumes to recruiters based on these information together with the user-user relationship (e.g., friends or following). The RecSys Challenge 2016~\cite{10.1145/2987538} is about the job recommendation problem.

Our person-job fit problem differs to the job (resp. resume) recommendation problem in two aspects.
First, job (resp. resume) recommendation is trying to predict and rank the jobs (resp. resume) based on the users' (resp. recruiters') preference; in person-job fit, given a candidate-job pair, we already know that the candidate has applied (i.e., shown interests) for the given job; the task is to predict whether the recruiter will offer him/her an interview or not. Second, in person-job fit, there is no social networking data or action data; instead, only resumes and job posts are available.

\subsection{Person-Job Fit}\label{sec:related-person-job}
Generally, good person-job fit algorithms should follow the manual review process, which checks the resume of the candidate and the job requirement to decide if they are matching. Understanding the semantics of the resume and job post is vital. Therefore, this task is highly related to natural language processing, especially text representation learning. Deep neural networks such as convolutional neural networks~\cite{kim-2014-convolutional} (CNN), recurrent neural networks~\cite{DBLP:conf/nips/SutskeverVL14} (RNN), attention models~\cite{DBLP:journals/corr/BahdanauCB14,DBLP:conf/nips/VaswaniSPUJGKP17} and BERT models~\cite{DBLP:conf/naacl/DevlinCLT19}, have made breakthrough in representation learning for text~\cite{DBLP:journals/corr/Goldberg15c}, including word, sentence and paragraph representation. Significant improvements are observed in search ranking where the similarity of the query and candidate documents are computed based on the learned representations~\cite{DBLP:conf/cikm/HuangHGDAH13,DBLP:conf/kdd/ShanHJWYM16}. 
These models can be adopted in our algorithm for learning representations of free text in resumes and job posts.

The most relevant works are \cite{acmtmis,DBLP:conf/cikm/RamanathIPHGOWK18,DBLP:conf/sigir/QinZXZJCX18,DBLP:conf/kdd/YanLSZZ019}. They all apply deep neural networks to learn representations for the resumes and jobs. 
Zhu et al. Chen et al.\cite{acmtmis} feed the embedding of each word in the resume and the job post into two CNN models respectively to extract their representations. Cosine similarity is applied against the extracted representations to calculate the matching score. 
Qin et al.\cite{DBLP:conf/sigir/QinZXZJCX18} use RNN models with attention hierarchically to learn word-level, single ability-aware and multiple ability-aware representations of the job post and resume respectively. The multiple ability-aware representations of the job post and the resume are combined and then fed into a binary classification sub-network. Yan et al. \cite{DBLP:conf/kdd/YanLSZZ019} analyze the interview history of the candidate and the job post to infer their preference. Memory networks are adopted to embed the preference in the representations of the candidate and the job post. Le et al.\cite{DBLP:conf/cikm/LeHSZ0019} define the intention of the candidate and the recruiter according to the actions including submitting a resume, accepting a resume and rejecting a resuming. Next, they train the representations of the job post and the resume to predict the intention rates and matching score simultaneously. Bian et al~\cite{bian-etal-2019-domain} learn the representations for jobs from some popular categories and then transfer the representations for jobs in other categories using domain adaptation algorithms.

Our method uses deep learning models to learn the job and resume representations as well. However, different to existing methods that learn the representations against only the free text in the resume and job post, our method additionally learns the representations of entities extracted from the whole resume and job post. The extracted entities are human readable, which help explain the matching results. Our method also exploits the actions as \cite{DBLP:conf/cikm/LeHSZ0019} to infer the preference (or intention) of candidates and recruiters; nonetheless, we learn the representations by accumulating all actions of each candidate and each recruiter instead of a single action. Our final representation of a candidate or job post is a fusion of the representations for the explicit and implicit intentions.

\section{Methodology}
\subsection{Problem Definition}\label{sec:problem}
We denote the candidate set as $C=\{c_i\}_{i=1}^m$ and the job post set as $P=\{p_j\}_{j=1}^n$. A set of historical applications (i.e., the training dataset) denoted as $H=\{(c_i, r_{ik}, p_j, t_{ik})\}$ is given, where $r_{ik}$ is the resume of candidate $c_i$ submitted to the job post $p_j$; $k$ indicates that this is the $k$-th resume (i.e., $k$-th application) submitted by $c_i$ \footnote{In this paper, we process the resume submitted in every application individually, although a candidate may submit the same resume for different job posts.}; $t_{ik}\in\{0, 1\}$ is the truth label for the matching result (1 for accept and 0 for reject). 
Figure.~\ref{fig:prob} illustrates the interactions between the candidates and job posts, where each edge represents a resume of a candidate submitted for a job. The same candidate may submit different resumes for different jobs.
The person-job fit task is to predict the matching result for a new application. It is possible that the candidate and the job post of this new application never appears in $H$. 
In the rest of this paper, we use $r, c, p, t$ respectively for the resume, candidate, job post  and truth label from an application when the context is clear.

\subsection{Overview}
Our overall idea is to learn an effective representation (i.e., feature vector) of the candidate and the job post, denoted as $f(c)$  and $g(p)$ respectively. The matching score is then calculated as $\sigma(f(c)^T g(p))$, i.e., the inner-product of the feature vectors normalized by sigmoid function. For the feature vector of the candidate (resp. job post), we do a late feature fusion, i.e., concatenation, of the feature vectors learned from two aspects. First, we learn a feature vector of the content of the job post to represent the recruiter's explicit intention, i.e., the job requirements,
denoted as $g_E(p)$ (resp. the resume $f_E(r)$ for the candidate's explicit skills). Second, we learn a feature vector to capture the implicit intention of the recruiter, denoted as $f_I(p)$, based on the previous applications he/she rejected and accepted. Similarly, we learn a feature vector to capture the implicit capabilities of the candidate denoted as $g_I(c)$ based on the applications he/she submitted, including accepted and rejected. Two neural network models are trained separately for the explicit and implicit features using the binary cross-entropy loss over the training tuples $H$,

\begin{eqnarray}
    L_E(c, r, p, t) &=& -t\ log\sigma(f_E(r)^T g_E(p)) \label{eq:loss_e}\\
    L_I(c, r, p, t) &=& -t\ log\sigma(f_I(c)^T g_I(p)) \label{eq:loss_i}
\end{eqnarray}

By fusing, i.e., concatenating, the feature vectors from the two models, we get $f(c) = [f_E(r); f_I(c)], g(p)=[g_E(p); g_I(p)]$, where $r$ is the resume of the candidate $c$ submitted for job post $p$. A threshold for the matching score $\sigma(f(c)^T g(p))$ is tuned using a validation dataset for making the final decision, i.e., accept or reject.

\subsection{Learning Explicit Features}\label{sec:explicit}
\begin{figure}
    \centering
    \includegraphics[width=0.48\textwidth]{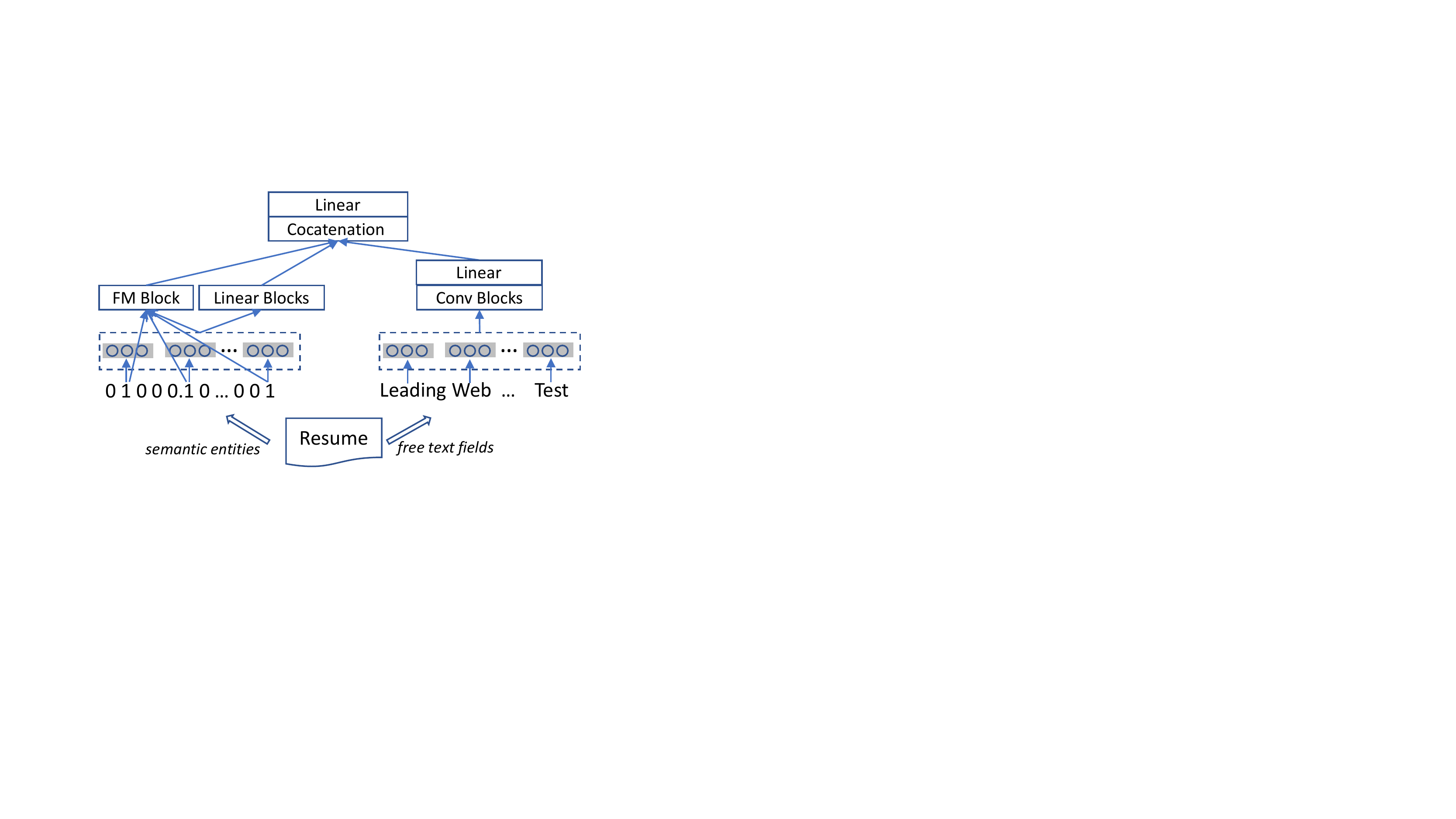}
    \caption{Model structure for learning explicit features.}
    \label{fig:explicit}
\end{figure}

In this subsection, we learn feature vectors directly for the resume and the job post from one application. The learned features represent the explicit information described in the resume and job post.
Both the job post and resume have many fields as illustrated in Figure~\ref{fig:prob}. For example, there are (simi-) structured fields (in blue text) such as education and skills in the resume, and job title and skill requirements in the job post. There are also free text fields (in gray text) such as working experience in the resume and job description in the job post. Most of existing works only learn features from the free text fields using deep learning techniques; that is to say, they skip the other fields, which in fact may contain valuable information. Towards this issue, we propose to extract semantic entities from the whole resume (and job post) using machine learning models and rules. The semantic entities are then embedded and fed into an adapted DeepFM ~\cite{DBLP:conf/ijcai/GuoTYLH17} to learn the correlations among the features, whose output feature vector is concatenated with the feature vector generated by a convolutional neural network against the free text fields. Figure~\ref{fig:explicit} shows the model structure. In this way, the final output feature vectors represent the resume and job post explicitly and comprehensively. Next, we shall introduce the algorithm to learn the features for resumes. The same algorithm is applied for job posts except for some hyper-parameters which are stated specifically.

\textbf{Semantic Entities}
Given a resume, we preprocess it to parse the content into a JSON format data structure, 
including education, working experience, skills, etc. For different fields, we apply different techniques to extract the semantic entities.
First, regular expressions and rules are applied to extract simple entities such as age, gender, location, etc. Some entities have different mentions (e.g., abbreviations) that are not easy to recognize using rules. Second, neural network classifiers are trained to extract these complex entities respectively, including university, major, position of the previous job, etc. For instance, we feed the text from the education field as the input to the model to predict the university index from a predefined university list. 
The BERT model~\cite{DBLP:conf/naacl/DevlinCLT19} is fine-tuned for this task.
Third, the rest entities are derived based on domain knowledge~\footnote{We created a knowledge base for this purpose. The knowledge base construction is out of the scope of this paper; therefore, we skip it.}
and the entities from the previous two steps. For example, we get the tier (e.g., top50) of a candidate's university by checking the university (from the second step) over an university ranking list, e.g., the QS ranking or USNews. After the three steps, we get a list of $s$ entities, among which some are categorical values, e.g., the university index, and some are real values, e.g., the duration of a candidate's previous job.

By converting every categorical entity into a one-hot representation and standardizing the real value entities, we get a sparse vector $\mathbf{x}$ of length $d_x$, 
which is fed into the adapted DeepFM model as shown in Figure~\ref{fig:explicit} (left). Different to the original DeepFM model that generates a single scalar value as the output of the FM Block, we produce a feature vector consisting of two parts: 1) $squeeze(\mathbf{w} * \mathbf{x})$, which multiplies the sparse vector  with a weight vector $\mathbf{w}\in R^{d_x}$ (to be trained) element-wisely and then removes the empty entries. Since there are $s$ entities in each resume, the output vector from $squeeze()$ has $s$ elements. This vector captures the first order feature of the entities. 2) $\sum_{i\neq j}\mathbf{V_i} * \mathbf{V_j} \mathbf{x}_i \mathbf{x}_j$, where $V_i\in R^{d_{fm}}$ is the dense embedding vector (to be trained), as shown in the shadow area, of the i-th element in $\mathbf{x}$.
The summation result is a vector of length $d_{fm}$. This vector represents the second-order features of the semantic entities. For the Linear Blocks in Figure~\ref{fig:explicit}, we follow the structure in the original DeepFM model. The input is a dense vector concatenated by all the embedding vectors of the $s$ entities. 
Finally, we concatenate the feature vectors from the FM block and Linear blocks as the output from the DeepFM side. 

\textbf{Free Text}
We follow \cite{acmtmis} to extract explicit features from the free text fields. The model is a convolutional neural network (CNN) with 2 2D convolution blocks and a linear layer. Each sentence from the free text fields is truncated and padded into a fixed length. A maximum sentence length is also set to get a fixed number of sentences via truncation and padding (at the sentence level). Therefore, the input to the CNN model is a matrix (one row per sentence) for each resume. The words are embedded through a dense embedding layer and the result (3D tensor) is fed into the convolution blocks. 

\textbf{Training}
The output feature vectors from the DeepFM model and the CNN model are concatenated and fed into a linear layer to produce the explicit feature vector of length $d_E$,
i.e., $f_E(r)$ and $g_E(p)$ for a resume and a job post respectively.
The whole model in Figure~\ref{fig:explicit} is trained end-to-end using back-propagation and mini-batch stochastic gradient descent algorithm (SGD) with the loss function defined in Equation~\ref{eq:loss_e}.

\subsection{Learning Implicit Features}\label{sec:implicit}
\textbf{Job Post}
A job post is written to describe the intention of the job recruiter; however, the text may not capture the intention comprehensively. For example, the recruiters may miss to mention some skills carelessly or dynamically change the expectation if the received resumes are of very high (or low) quality.
Nonetheless, previously accepted and rejected resumes can help infer these implicit intentions. Therefore, we exploit the application history to learn the implicit features for each job post.

\begin{figure}
    \centering
    \includegraphics[width=0.5\textwidth]{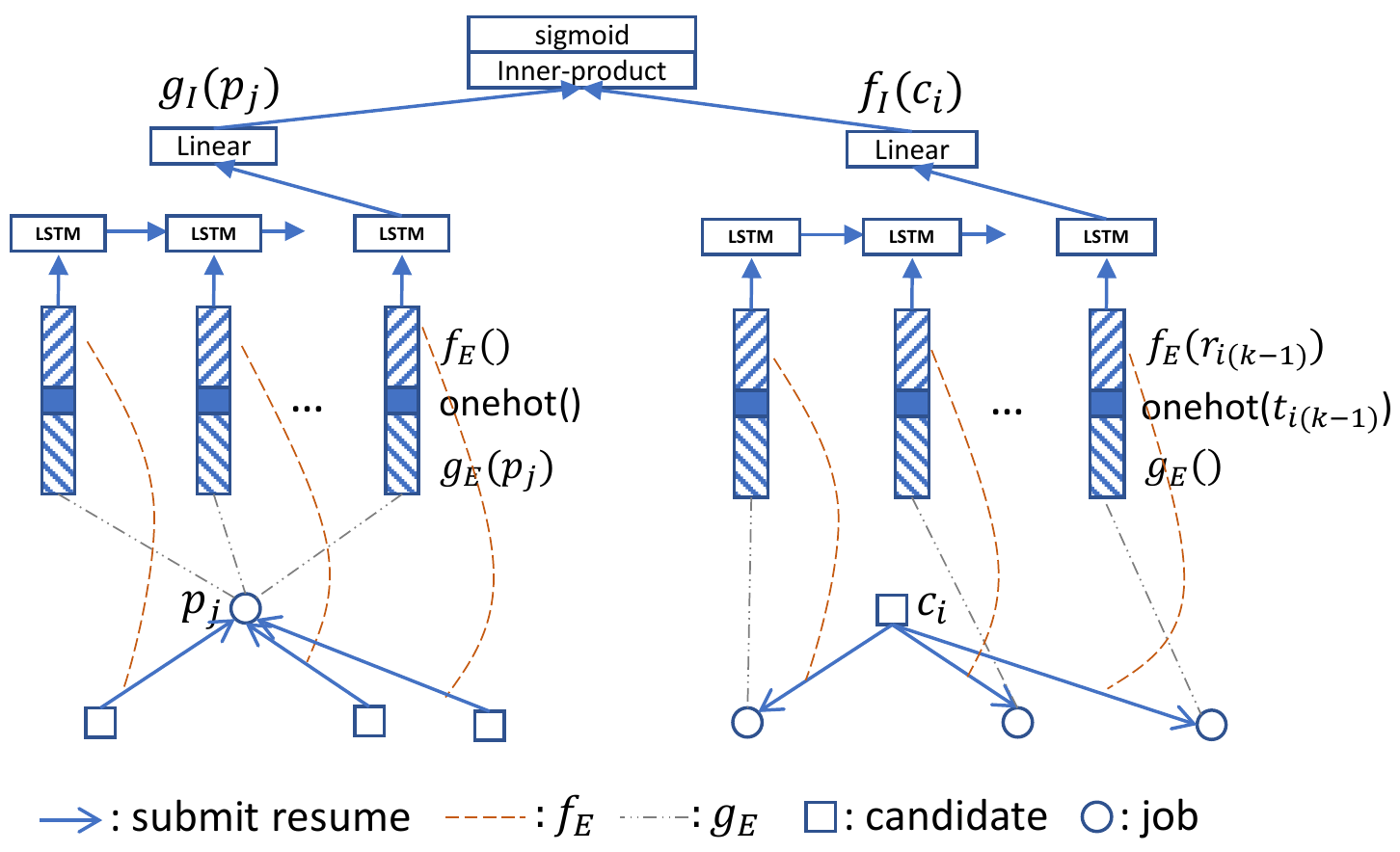}
    \caption{Model structure for learning implicit features.}
    \label{fig:implicit}
\end{figure}

Given a job post $p_j$ from an application, we extract all applications to $p_j$ before this one from $H$ as $H(p_j)=\{(c_i, r_{ik}, p_j, t_{ik})\}$. A LSTM model\footnote{We leave the testing of other advanced RNN models, e.g., Bi-directional LSTM, as a future work.} is applied to learn the features from $H(p_j)$, as shown in Figure~\ref{fig:implicit} (left side). Specifically, we propose a novel input encoding scheme for each tuple in $H(p_j)$. For each $(c_i, r_{ik}, p_j, t_{ik})\in H(p_j)$, we concatenate the explicit feature vectors as $[f_E(r_{ik})$; $onehot(t_{ik})$; $g_E(p_j)]$ $\in R^{2d_E+2}$, where onehot($t_{ik}$) is the onehot representation of the result, i.e, $01$ for rejecting or $10$ for accepting. By feeding the explicit features of the resumes and the current job post as well as the decision (i.e., accept or reject), we expect the LSTM model to learn the features for representing the implicit intention of $p_j$.  These encoded vectors are fed into the LSTM model according to time when they are reviewed by the recruiter. The last hidden state vector of the LSTM model is transformed by a linear layer with $d_I$ neurons, whose output is assigned to $g_I(p_j)$.

\textbf{Candidate}
Since a candidate may change the resume over time and submit different versions for different job posts, it is more reasonable to learn the implicit intention of the candidate instead of every resume. The same model structure is applied to learn the implicit features for each candidate as shown in Figure~\ref{fig:implicit} (right side). Note that for each job $p_j$ applied by candidate $c_i$, it is encoded into the input to the LSTM as $[f_E(r_{ik});$ $onehot(t_{ik});$ $g_I(p_j)]\in R^{2d_E+2}$, where $r_{ik}$ is the resume submitted for $p_j$. The last hidden vector of the LSTM model is transformed by a linear layer with $d_I$ neurons to generate $f_I(c_i)$.

\textbf{Training}
The whole model in Figure~\ref{fig:implicit} is trained end-to-end by back-propagation and SGD  with Equation~\ref{eq:loss_i} as the loss function. For each training tuple (i.e., application) $(c_i, r_{ik}, p_j, t_{ik})\in H$, we collect all previously applied jobs for $c_i$ and all previously received resumes for $p_j$ to construct the inputs for the two LSTM models respectively. To facilitate fast training, we truncate or pad the inputs to have a fixed length of 20 (left) and 5 (right) respectively. The numbers are derived with reference to the average number of resumes per job and average number of jobs per candidate. If padding is applied, the onehot vector for the decision is $00$ (for unknown).
For the case where there is no historical data for $p_j$ (or $c_i$), all the input resumes are generated via padding. 
Note such cases take a small portion of all tuples as it happens only when the job post or the candidate is new to the system. 
Although the output does not carry any extra (intention) information, we still concatenate it with the explicit features to make all job posts and candidates have the feature vectors of the same length, i.e., $d_E+d_I$. With that, we can compute a matching score between any job post and candidate.

\section{Experiment}

\subsection{Experiment setup}

\textbf{Dataset} Both the resumes and job posts are sensitive data. As far as we know, there is no public data available for the person-job fit problem. Existing papers use private data provided by commercial recruitment platforms. In our experiment, we use the real data collected by our recruitment platform from January 2019 to October 2019. 
In total, there about 13K (K=1000) job posts, 580K candidates and 1.3 million resumes. 
Each application is a data sample (i.e., a tuple defined in Section~\ref{sec:problem}). We order them according to the time when the recruiter reviews the application, which is in accordance with the order to the LSTM model.
The last 100K samples are used for testing and the second last 100K samples are used for validation; the rest (about 1.1 million) is used for training.


\textbf{Baselines}
We compare our feature fusion solution, called PJFFF (Person-Job Fit based on Feature Fusion) with the following baseline methods, including traditional machine learning models, a collaborative filtering algorithm and three deep learning based methods.
More details of the last three methods are introduced in Section~\ref{sec:related-person-job}.
\begin{itemize}
    \item Logistic regression (LR), Factorized Machine (FM), Multi-layer Perceptron (MLP) and LightGBM~\cite{DBLP:conf/nips/KeMFWCMYL17}. For each resume and job post pair (from one application), we feed the sparse vector $\mathbf{x}$ from Section~\ref{sec:explicit} as the input to these models.
    
    \item LightGCN~\cite{he2020lightgcn} proposes a graph convolutional neural network to implement collaborative filtering. The embedding vectors of the candidate nodes and job nodes are learned and compared (inner-product) for recommendation.
    
    \item PJFNN~\cite{acmtmis} uses two CNN models to learn the features of the free text in the resume and job post respectively. The structure is the same as that for learning the explicit features of the free text in Figure~\ref{fig:explicit} (right side).
    
    \item APJFNN \cite{DBLP:conf/sigir/QinZXZJCX18} applies attention modelling recursively to learn ability-aware representations for the resume and job post. 
    
    \item JRMPM\cite{DBLP:conf/kdd/YanLSZZ019} also exploits the application history. It uses a memory network to embed the preference of the candidate and the job recruiter into their representations.
\end{itemize}
LR, FM, MLP and our model are trained using PyTorch; We use the open-source implementation of LightGBM and LightGCN. The codes of the PJFNN, APJFNN and JRMPM are kindly shared by the authors of \cite{bian-etal-2019-domain,DBLP:conf/kdd/YanLSZZ019}.

\textbf{Evaluation Metrics}
Following previous works~\cite{DBLP:conf/kdd/YanLSZZ019,DBLP:conf/sigir/QinZXZJCX18}, we evaluate the accuracy, AUC, precision, recall and F1 of each method. We tune the threshold for generating the decision (accept or reject) over the validation dataset to maximize the F1 metric. The same threshold is applied for computing the accuracy metric. According to the business requirement, we tune another threshold to make the recall at 0.8 and then compute the precision@recall=0.8.


\textbf{Hyper-parameter Setting}. 
In our experiment, we extract $s=264$ entities for each resume and $s=57$ entities for each job post. The length of the sparse feature vector $d_x$ is about 37K for a resume and 1.6K for a job post respectively. For the adapted DeepFM model in Figure~\ref{fig:explicit}, we set embedding vector size as $d_{fm}=7$ and the size of the last linear layer as $d_E=128$.
The CNN model in Figure~\ref{fig:explicit} has the same structure as that in \cite{acmtmis}. The default size of the hidden state vector and the linear layer in the LSTM models is set to 64. Hyper-parameter sensitivity analysis is conducted for them in Section~\ref{sec:hyper}.
To train the models in Figure~\ref{fig:explicit} and Figure~\ref{fig:implicit}, we use mini-batch Adam~\cite{kingma2017method} as the optimizer with batchsize 512. The learning rate is set to 0.005 and 0.001 for training the models in Section~\ref{sec:explicit} and Section~\ref{sec:implicit} respectively. The weight decay (coefficient or L2 regularization) is  $10^{-5}$ and $10^{-4}$ for the two models respectively. 

\subsection{Performance Comparison}

The performance metrics of all methods are listed in Table~\ref{tab:perf}. We can see that our solution PJFFF outperforms all baseline methods across all evaluation metrics with a large margin. PJFNN, APJFNN and JRMPM are the most relevant methods to our solution. However, they fail to read and process the non-free text fields in resumes and job posts, which actually may contain valuable information to capture the explicit intention of the candidates and recruiters. Moreover, PJFNN and APJFNN ignore the historical applications of each candidate and each job post. In contrast, JRMPM and PJFFF both utilize the historical information to infer the implicit intention. Therefore, they get better performance. 

LightGBM is the second best method, which even outperforms the recently proposed deep-learning-based methods, namely, PJFNN, APJFNN and JRMPM. In fact, LR, FM and DNN are comparable to the deep-learning-based methods. We notice that in APJFNN\cite{DBLP:conf/sigir/QinZXZJCX18} and JRMPM\cite{DBLP:conf/kdd/YanLSZZ019} papers,  these methods' performance is much worse than APJFNN and JRMRM. We think the major reason is that we are using different input feature vectors. In APJFNN and JRMPM papers, they average the embeddings of all words from the free text fields of a resume and a job post respectively as the inputs to GBDT. In our experiment, we feed the sparse entity feature vector into these model. The result shows that our entity feature vector is effective in representing the resume and job post.

LightGCN has the worst performance. Although it has shown to be effective than other collaborative filtering approaches for recommendation problems, LightGCN is not a good solution for person-job fit. We think this is mainly because LightGCN does not use the content of the resume and job post at all. It learns the embedding vectors of each resume and each job post merely based on the interactions among the resume nodes and job post nodes in the graph. Moreover, some nodes do not have connections to others. In other words, for new candidates and job posts, LightGCN has the cold starting issue, which is a common challenge for collaborative filtering methods. For our method, although we cannot get any implicit intention information for such cases, we still have the explicit features.

\begin{table}[]
    \centering
    \caption{Performance comparison for person-job fit.}
    \begin{tabular}{|l|l|l|l|l|}
    \hline 
         Method& AUC  & Accuracy &Precision & F1 \\ 
         &(\%) &(\%) & @Recall=0.8 (\%)&(\%) \\ \hline 
         LR  & 88.6 & 87.6 & 48.5 & 50.2 \\
         FM & 90.0  & 88.7 & 51.2 & 60.9 \\
         DNN & 91.2 & 89.4 & 55.9 & 64.9 \\
         LightGBM& 91.7 & 90.2 & 56.9 & 67.6\\
         LightGCN & 64.1 & 87.5 & 12.7 & 36.5\\ \hline
         PJFNN & 89.6 & 89.3 & 46.0 & 64.2 \\
         APJFNN & 90.4& 90.8& 48.7& 68.0 \\
         JRMPM & 91.3 & 88.7 & 51.8 & 66.7\\ \hline 
         PJFFF & \textbf{95.3}& \textbf{92.9}& \textbf{73.3}& \textbf{77.1}\\ \hline 
    \end{tabular}
    \label{tab:perf}
\end{table}

\subsection{Ablation Study}
Our PJFFF fuses two sets of features, namely explicit and implicit features. The explicit features are generated by the adapted DeepFM and CNN. Consequently, it has multiple variations by combining these components differently. In this section, we evaluate four variations to study the contribution of each component. 
\begin{enumerate}
\item $f_E\&g_E (entity)$ This variation uses the adapted DeepFM model only to learn the explicit features, i.e., the left part of Figure~\ref{fig:explicit}. The free text data is ignored.

\item $f_E\&g_E (both)$ This variation learns the explicit features using both DeepFM and CNN, i.e., Figure~\ref{fig:explicit}. No implicit features are learned.

\item $f\&g (entity)$ This variation fuses the explicit features generated by DeepFM and the implicit features without using the free text data.

\item $f\&g (both)$ This is the complete PJFFF model which fuses the implicit features and explicit features over the whole resume and job post.

\end{enumerate}

\begin{table}[]
    \centering
    \caption{Ablation study of our PJFFN.}
    \begin{tabular}{|l|l|l|l|l|}
    \hline 
         Method& AUC & Accuracy &Precision & F1 \\ 
         & (\%)& (\%)& @Recall=0.8(\%)&(\%) \\ \hline
         $f_E \& g_E$ (entity) & 91.4 & 89.6 & 56.4 & 65.7\\
         $f_E \& g_E$ (both) & 94.2 & 91.0& 65.6& 73.1\\ 
         $f \& g$ (entity) & 93.1 & 91.5 & 63.3 & 71.3\\  
         $f \& g$ (both) & 95.3 & 92.9& 73.3 & 77.1 \\ \hline 
    \end{tabular}
    \label{tab:ablation}
\end{table}


The ablation study results are listed in Table~\ref{tab:ablation}. First, the $f_E\&g_E ($ $entity)$ variation has the worst performance. However, although it only learns the explicit features using DeepFM over the semantic entities, the performance is comparable to other baseline methods in Table~\ref{tab:perf}. The result confirms that the semantic entities carry valuable information. Second, by comparing the performance of $f_E\&g_E (both)$ and $f_E\&g_E (entity)$, we can see there is a significant improvement. This is due to the contribution of the CNN model over the free text fields. Similar improvement is also observed by comparing $f \& g (entity)$ and $f \& g (both)$. Third, when we compare the rows for $f_E\&g_E (entity)$ and $f \& g (entity)$ (resp. $f_E\&g_E (both)$ and $f \& g (both)$), the performance is improved. It indicates that the implicit features make positive contributions. To conclude, all components involved in PJFFF contributes positively to the final performance. By combining them together, we get the best model.

\subsection{Hyper-parameter Sensitivity Study}\label{sec:hyper}

\begin{figure}[h]
    \begin{subfigure}{0.23\textwidth}
    \centering
    \includegraphics[width=\textwidth]{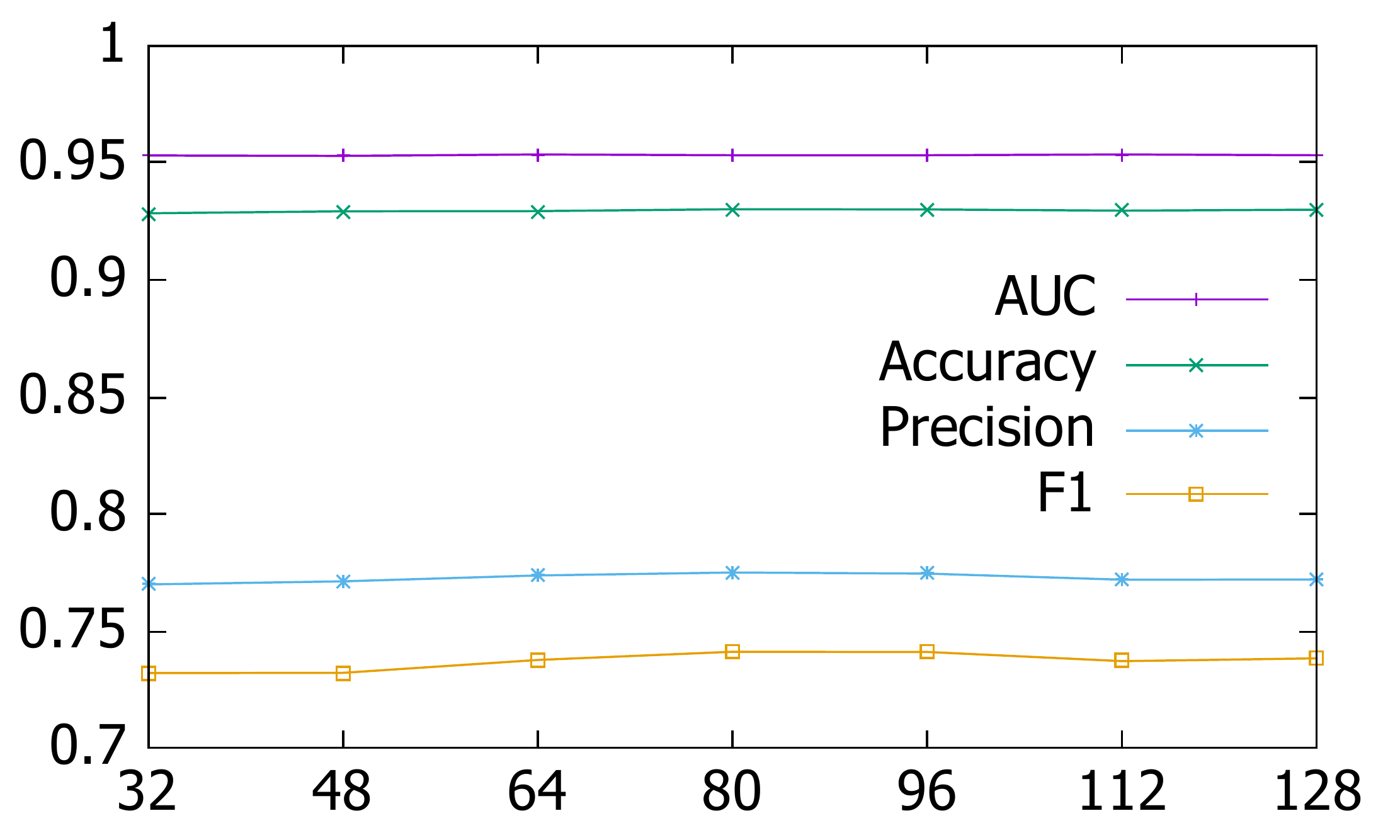}
    \caption{Size of the linear layer.}
    \label{fig:hidsize}
    \end{subfigure}
    \hfill
    \begin{subfigure}{0.23\textwidth}
    \centering
    \includegraphics[width=\textwidth]{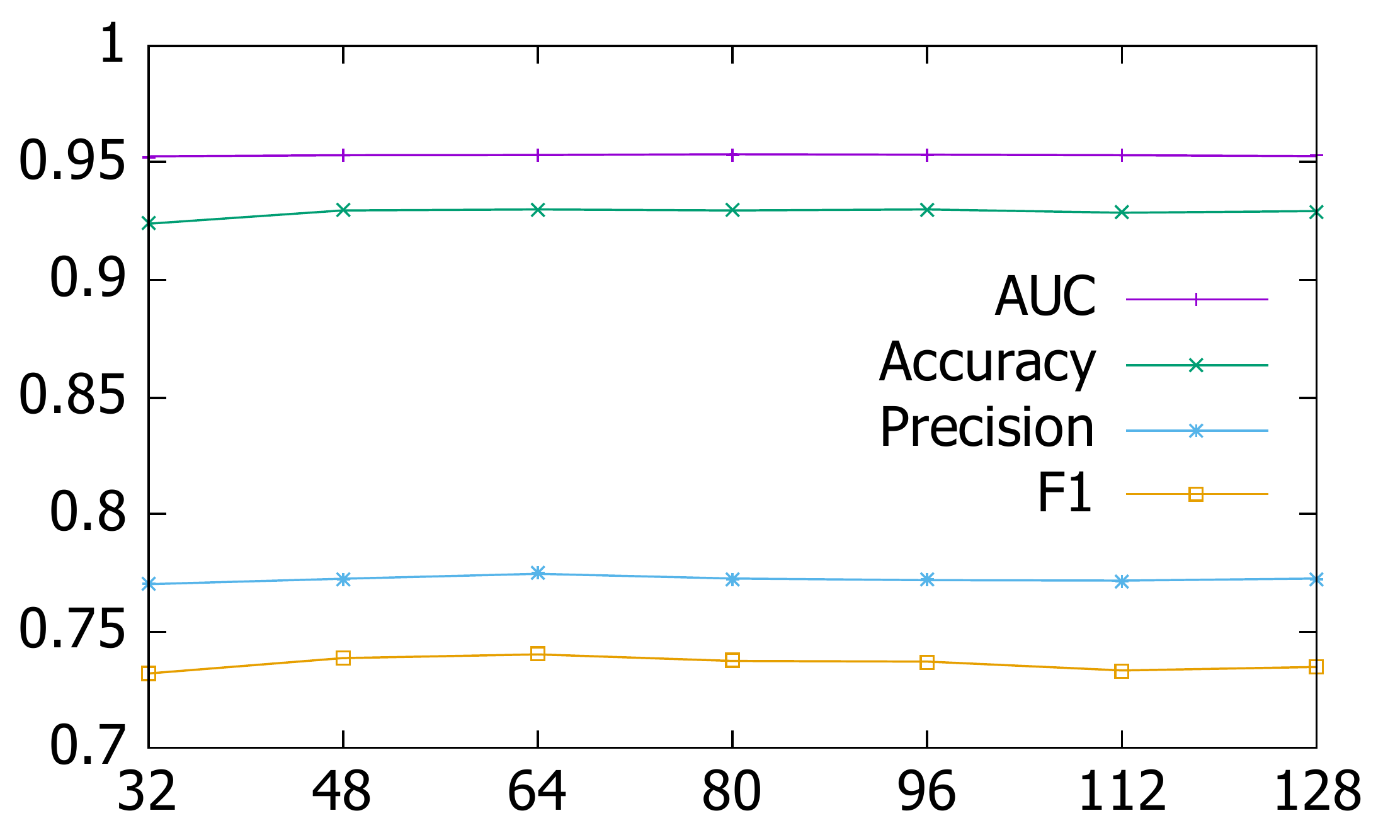}
    \caption{Size of the hidden state.}
    \label{fig:lsize}
    \end{subfigure}
    \caption{Sensitivity analysis of hyper-parameters.}
    \label{fig:sensititivy}
\end{figure}

In this section, we study the sensitivity of two hyper-parameters of PJFFF pertaining to the model structure, namely the hidden state size of the LSTM layer and the last linear layer size in Figure~\ref{fig:implicit}. Other model structure hyper-parameters are mainly set following existing papers; and the training hyper-parameters are tuned over the validation dataset. Figure~\ref{fig:sensititivy} shows the results when we vary the size. We can see that the performance metrics are almost on horizontal lines, indicating that PJFFF is not sensitive to the two hyper-parameters.

\begin{figure*}[t]
\centering
\includegraphics[width=0.95\textwidth]{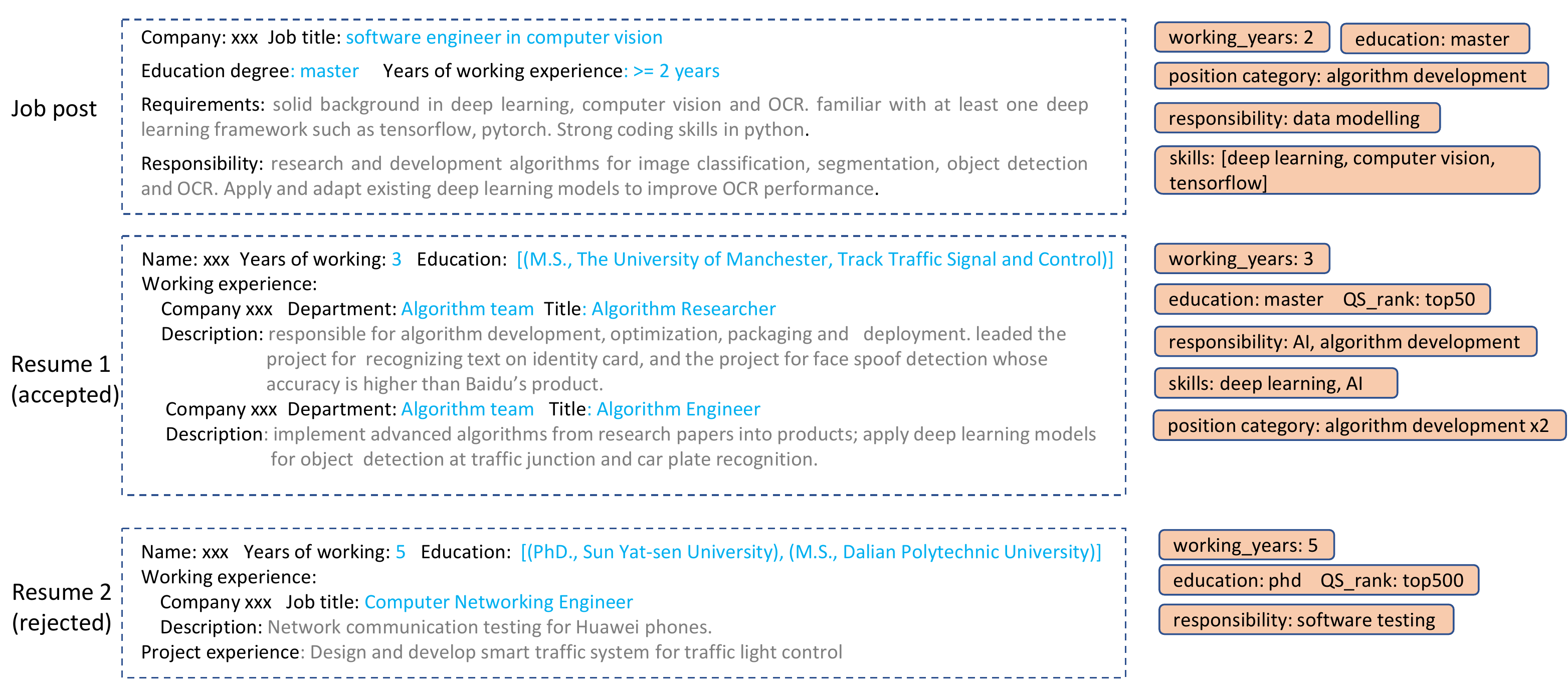}
\caption{From top to bottom: Job post, accepted resume and rejected resume.}
\label{fig:case}
\end{figure*}

\subsection{Case Study}
Case studies are conducted to inspect and verify our solution from another perspective. Figure~\ref{fig:case} shows two application cases for the same job. 
The left side illustrates the job post and resumes~\footnote{We reconstruct the job post and resumes from the parsed JSON files and present a subset of the fields for better illustration.}. The right side displays some extracted semantic entities. The matching score generated for Resume 1 and Resume 2 is 0.78 and 0.04 respectively. We can see that our solution is able to rank the accepted resume and rejected resume correctly. To understand why the two resumes get different scores, we can refer to the semantic entities, which clearly indicate the requirement of this job, and the experience and capability of the candidates. We can see that the entities of Resume 1 match those of the job post well. For example, the candidate worked on algorithm development for two jobs, and he/she has the deep learning and AI skills which are required by the job post. However, for Resume 2, the semantic entities are less relevant to those of the job post. To conclude, these semantic entities are informative for interpreting the result generated by our method. For example, human resource (HR) managers who do not have the expertise knowledge of the position can refer to the semantic entities to understand the model's decision.

\subsection{Online Deployment}
Our solution is developed in 2 stages, where the first stage learns the explicit features using the semantic entities as the input. For this part, we have deployed it for online serving for  1 year. Some special optimization and engineering work is done to improve the efficiency and reduce the cost for large-scale deployment. In specific, considering that GPUs are expensive, we deploy our method on CPU machines. To improve the efficiency, we re-implement all pre-processing steps and semantic entity extraction steps using the Go language, which reduces the inference time per sample by 50\%. We also applied model distillation~\cite{DBLP:journals/corr/abs-1903-12136} to replace the BERT model for extracting the complex semantic entities. 
After that, the inference time is reduced by another 50\% and is 0.2-0.3 second per sample. The second stage adds the rest components introduced in this paper. These additional components increase the time complexity. We are optimizing the efficiency and will deploy them later.

\section{Conclusion}
In this paper, we have introduced a feature fusion based method, called PJFFF, for the person-job fit problem. Our method differs to existing solutions in two aspects. First, instead of purely relying on deep learning models to process the free text fields in a resume and job post, PJFFF extract semantic entities from the whole resume and job post in order to learn comprehensive explicit features. Second, we propose a new scheme to encode the historical application data of each candidate and each job post. They are fed into LSTM models to learn implicit intentions of the candidate and recruiters respectively. Finally, the two sets of features are fused together via concatenation to produce an effective representation. Experimental study confirms the superiority of our method over existing methods and verifies the contributions from each component of PJFFF.


\bibliographystyle{ACM-Reference-Format}
\bibliography{main}


\begin{thebibliography}{00}


\ifx \showCODEN    \undefined \def \showCODEN     #1{\unskip}     \fi
\ifx \showDOI      \undefined \def \showDOI       #1{#1}\fi
\ifx \showISBNx    \undefined \def \showISBNx     #1{\unskip}     \fi
\ifx \showISBNxiii \undefined \def \showISBNxiii  #1{\unskip}     \fi
\ifx \showISSN     \undefined \def \showISSN      #1{\unskip}     \fi
\ifx \showLCCN     \undefined \def \showLCCN      #1{\unskip}     \fi
\ifx \shownote     \undefined \def \shownote      #1{#1}          \fi
\ifx \showarticletitle \undefined \def \showarticletitle #1{#1}   \fi
\ifx \showURL      \undefined \def \showURL       {\relax}        \fi
\providecommand\bibfield[2]{#2}
\providecommand\bibinfo[2]{#2}
\providecommand\natexlab[1]{#1}
\providecommand\showeprint[2][]{arXiv:#2}

\bibitem[\protect\citeauthoryear{??}{10.}{2016}]%
        {10.1145/2987538}
 \bibinfo{year}{2016}\natexlab{}.
\newblock \bibinfo{booktitle}{{\em RecSys Challenge ’16: Proceedings of the
  Recommender Systems Challenge}}. \bibinfo{publisher}{Association for
  Computing Machinery}, \bibinfo{address}{New York, NY, USA}.
\newblock
\showISBNx{9781450348010}


\bibitem[\protect\citeauthoryear{Bahdanau, Cho, and Bengio}{Bahdanau
  et~al\mbox{.}}{2015}]%
        {DBLP:journals/corr/BahdanauCB14}
\bibfield{author}{\bibinfo{person}{Dzmitry Bahdanau},
  \bibinfo{person}{Kyunghyun Cho}, {and} \bibinfo{person}{Yoshua Bengio}.}
  \bibinfo{year}{2015}\natexlab{}.
\newblock \showarticletitle{Neural Machine Translation by Jointly Learning to
  Align and Translate}. In \bibinfo{booktitle}{{\em {ICLR}}},
  \bibfield{editor}{\bibinfo{person}{Yoshua Bengio} {and} \bibinfo{person}{Yann
  LeCun}} (Eds.).
\newblock


\bibitem[\protect\citeauthoryear{{Bau}, {Zhou}, {Khosla}, {Oliva}, and
  {Torralba}}{{Bau} et~al\mbox{.}}{2017}]%
        {8099837}
\bibfield{author}{\bibinfo{person}{D. {Bau}}, \bibinfo{person}{B. {Zhou}},
  \bibinfo{person}{A. {Khosla}}, \bibinfo{person}{A. {Oliva}}, {and}
  \bibinfo{person}{A. {Torralba}}.} \bibinfo{year}{2017}\natexlab{}.
\newblock \showarticletitle{Network Dissection: Quantifying Interpretability of
  Deep Visual Representations}. In \bibinfo{booktitle}{{\em {CVPR}}}.
  \bibinfo{pages}{3319--3327}.
\newblock


\bibitem[\protect\citeauthoryear{Bian, Zhao, Song, Zhang, and Wen}{Bian
  et~al\mbox{.}}{2019}]%
        {bian-etal-2019-domain}
\bibfield{author}{\bibinfo{person}{Shuqing Bian}, \bibinfo{person}{Wayne~Xin
  Zhao}, \bibinfo{person}{Yang Song}, \bibinfo{person}{Tao Zhang}, {and}
  \bibinfo{person}{Ji-Rong Wen}.} \bibinfo{year}{2019}\natexlab{}.
\newblock \showarticletitle{Domain Adaptation for Person-Job Fit with
  Transferable Deep Global Match Network}. In \bibinfo{booktitle}{{\em
  EMNLP-IJCNLP}}. \bibinfo{publisher}{Association for Computational
  Linguistics}, \bibinfo{address}{Hong Kong, China},
  \bibinfo{pages}{4810--4820}.
\newblock
\showDOI{%
\url{https://doi.org/10.18653/v1/D19-1487}}


\bibitem[\protect\citeauthoryear{Cheng, Xie, Chen, Agrawal, Choudhary, and
  Guo}{Cheng et~al\mbox{.}}{2013}]%
        {DBLP:conf/kdd/ChengXCACG13}
\bibfield{author}{\bibinfo{person}{Yu Cheng}, \bibinfo{person}{Yusheng Xie},
  \bibinfo{person}{Zhengzhang Chen}, \bibinfo{person}{Ankit Agrawal},
  \bibinfo{person}{Alok~N. Choudhary}, {and} \bibinfo{person}{Songtao Guo}.}
  \bibinfo{year}{2013}\natexlab{}.
\newblock \showarticletitle{JobMiner: a real-time system for mining job-related
  patterns from social media}. In \bibinfo{booktitle}{{\em {SIGKDD}}}.
  \bibinfo{publisher}{{ACM}}, \bibinfo{pages}{1450--1453}.
\newblock
\showDOI{%
\url{https://doi.org/10.1145/2487575.2487704}}


\bibitem[\protect\citeauthoryear{Devlin, Chang, Lee, and Toutanova}{Devlin
  et~al\mbox{.}}{2019}]%
        {DBLP:conf/naacl/DevlinCLT19}
\bibfield{author}{\bibinfo{person}{Jacob Devlin}, \bibinfo{person}{Ming{-}Wei
  Chang}, \bibinfo{person}{Kenton Lee}, {and} \bibinfo{person}{Kristina
  Toutanova}.} \bibinfo{year}{2019}\natexlab{}.
\newblock \showarticletitle{{BERT:} Pre-training of Deep Bidirectional
  Transformers for Language Understanding}. In \bibinfo{booktitle}{{\em
  {NAACL-HLT}}}. \bibinfo{publisher}{Association for Computational
  Linguistics}, \bibinfo{pages}{4171--4186}.
\newblock
\showDOI{%
\url{https://doi.org/10.18653/v1/n19-1423}}


\bibitem[\protect\citeauthoryear{Diaby, Viennet, and Launay}{Diaby
  et~al\mbox{.}}{2013a}]%
        {DBLP:conf/asunam/DiabyVL13}
\bibfield{author}{\bibinfo{person}{Mamadou Diaby}, \bibinfo{person}{Emmanuel
  Viennet}, {and} \bibinfo{person}{Tristan Launay}.}
  \bibinfo{year}{2013}\natexlab{a}.
\newblock \showarticletitle{Toward the next generation of recruitment tools: an
  online social network-based job recommender system}. In
  \bibinfo{booktitle}{{\em {ASONAM}}},
  \bibfield{editor}{\bibinfo{person}{Jon~G. Rokne} {and}
  \bibinfo{person}{Christos Faloutsos}} (Eds.). \bibinfo{publisher}{{ACM}},
  \bibinfo{pages}{821--828}.
\newblock
\showDOI{%
\url{https://doi.org/10.1145/2492517.2500266}}


\bibitem[\protect\citeauthoryear{Diaby, Viennet, and Launay}{Diaby
  et~al\mbox{.}}{2013b}]%
        {10.1145/2492517.2500266}
\bibfield{author}{\bibinfo{person}{Mamadou Diaby}, \bibinfo{person}{Emmanuel
  Viennet}, {and} \bibinfo{person}{Tristan Launay}.}
  \bibinfo{year}{2013}\natexlab{b}.
\newblock \showarticletitle{Toward the next Generation of Recruitment Tools: An
  Online Social Network-Based Job Recommender System}. In
  \bibinfo{booktitle}{{\em {ASONAM}}}. \bibinfo{publisher}{Association for
  Computing Machinery}, \bibinfo{address}{New York, NY, USA},
  \bibinfo{pages}{821–828}.
\newblock
\showISBNx{9781450322409}
\showDOI{%
\url{https://doi.org/10.1145/2492517.2500266}}


\bibitem[\protect\citeauthoryear{Goldberg}{Goldberg}{2015}]%
        {DBLP:journals/corr/Goldberg15c}
\bibfield{author}{\bibinfo{person}{Yoav Goldberg}.}
  \bibinfo{year}{2015}\natexlab{}.
\newblock \showarticletitle{A Primer on Neural Network Models for Natural
  Language Processing}.
\newblock \bibinfo{journal}{{\em CoRR\/}}  \bibinfo{volume}{abs/1510.00726}
  (\bibinfo{year}{2015}).
\newblock
\showeprint[arxiv]{1510.00726}
\showURL{%
\url{http://arxiv.org/abs/1510.00726}}


\bibitem[\protect\citeauthoryear{Guo, Tang, Ye, Li, and He}{Guo
  et~al\mbox{.}}{2017}]%
        {DBLP:conf/ijcai/GuoTYLH17}
\bibfield{author}{\bibinfo{person}{Huifeng Guo}, \bibinfo{person}{Ruiming
  Tang}, \bibinfo{person}{Yunming Ye}, \bibinfo{person}{Zhenguo Li}, {and}
  \bibinfo{person}{Xiuqiang He}.} \bibinfo{year}{2017}\natexlab{}.
\newblock \showarticletitle{DeepFM: {A} Factorization-Machine based Neural
  Network for {CTR} Prediction}. In \bibinfo{booktitle}{{\em {IJCAI}}},
  \bibfield{editor}{\bibinfo{person}{Carles Sierra}} (Ed.).
  \bibinfo{publisher}{ijcai.org}, \bibinfo{pages}{1725--1731}.
\newblock
\showDOI{%
\url{https://doi.org/10.24963/ijcai.2017/239}}


\bibitem[\protect\citeauthoryear{He, Deng, Wang, Li, Zhang, and Wang}{He
  et~al\mbox{.}}{2020}]%
        {he2020lightgcn}
\bibfield{author}{\bibinfo{person}{Xiangnan He}, \bibinfo{person}{Kuan Deng},
  \bibinfo{person}{Xiang Wang}, \bibinfo{person}{Yan Li},
  \bibinfo{person}{Yongdong Zhang}, {and} \bibinfo{person}{Meng Wang}.}
  \bibinfo{year}{2020}\natexlab{}.
\newblock \bibinfo{title}{LightGCN: Simplifying and Powering Graph Convolution
  Network for Recommendation}.
\newblock   (\bibinfo{year}{2020}).
\newblock
\showeprint[arxiv]{cs.IR/2002.02126}


\bibitem[\protect\citeauthoryear{Huang, He, Gao, Deng, Acero, and Heck}{Huang
  et~al\mbox{.}}{2013}]%
        {DBLP:conf/cikm/HuangHGDAH13}
\bibfield{author}{\bibinfo{person}{Po{-}Sen Huang}, \bibinfo{person}{Xiaodong
  He}, \bibinfo{person}{Jianfeng Gao}, \bibinfo{person}{Li Deng},
  \bibinfo{person}{Alex Acero}, {and} \bibinfo{person}{Larry~P. Heck}.}
  \bibinfo{year}{2013}\natexlab{}.
\newblock \showarticletitle{Learning deep structured semantic models for web
  search using clickthrough data}. In \bibinfo{booktitle}{{\em {CIKM}}},
  \bibfield{editor}{\bibinfo{person}{Qi~He}, \bibinfo{person}{Arun Iyengar},
  \bibinfo{person}{Wolfgang Nejdl}, \bibinfo{person}{Jian Pei}, {and}
  \bibinfo{person}{Rajeev Rastogi}} (Eds.). \bibinfo{publisher}{{ACM}},
  \bibinfo{pages}{2333--2338}.
\newblock
\showDOI{%
\url{https://doi.org/10.1145/2505515.2505665}}


\bibitem[\protect\citeauthoryear{Ke, Meng, Finley, Wang, Chen, Ma, Ye, and
  Liu}{Ke et~al\mbox{.}}{2017}]%
        {DBLP:conf/nips/KeMFWCMYL17}
\bibfield{author}{\bibinfo{person}{Guolin Ke}, \bibinfo{person}{Qi Meng},
  \bibinfo{person}{Thomas Finley}, \bibinfo{person}{Taifeng Wang},
  \bibinfo{person}{Wei Chen}, \bibinfo{person}{Weidong Ma},
  \bibinfo{person}{Qiwei Ye}, {and} \bibinfo{person}{Tie{-}Yan Liu}.}
  \bibinfo{year}{2017}\natexlab{}.
\newblock \showarticletitle{LightGBM: {A} Highly Efficient Gradient Boosting
  Decision Tree}. In \bibinfo{booktitle}{{\em {NIPS}}}.
  \bibinfo{pages}{3146--3154}.
\newblock


\bibitem[\protect\citeauthoryear{Kim}{Kim}{2014}]%
        {kim-2014-convolutional}
\bibfield{author}{\bibinfo{person}{Yoon Kim}.} \bibinfo{year}{2014}\natexlab{}.
\newblock \showarticletitle{Convolutional Neural Networks for Sentence
  Classification}. In \bibinfo{booktitle}{{\em {EMNLP}}}.
  \bibinfo{publisher}{Association for Computational Linguistics},
  \bibinfo{address}{Doha, Qatar}, \bibinfo{pages}{1746--1751}.
\newblock
\showDOI{%
\url{https://doi.org/10.3115/v1/D14-1181}}


\bibitem[\protect\citeauthoryear{Kingma and Adam}{Kingma and Adam}{2017}]%
        {kingma2017method}
\bibfield{author}{\bibinfo{person}{DP Kingma} {and} \bibinfo{person}{Ba~J
  Adam}.} \bibinfo{year}{2017}\natexlab{}.
\newblock \showarticletitle{A method for stochastic optimization. cornell
  university library}.
\newblock \bibinfo{journal}{{\em arXiv preprint arXiv:1412.6980\/}}
  (\bibinfo{year}{2017}).
\newblock


\bibitem[\protect\citeauthoryear{Koh and Liang}{Koh and Liang}{2017}]%
        {DBLP:conf/icml/KohL17}
\bibfield{author}{\bibinfo{person}{Pang~Wei Koh} {and} \bibinfo{person}{Percy
  Liang}.} \bibinfo{year}{2017}\natexlab{}.
\newblock \showarticletitle{Understanding Black-box Predictions via Influence
  Functions}. In \bibinfo{booktitle}{{\em {ICML}}} {\em
  (\bibinfo{series}{Proceedings of Machine Learning Research})},
  \bibfield{editor}{\bibinfo{person}{Doina Precup} {and}
  \bibinfo{person}{Yee~Whye Teh}} (Eds.), Vol.~\bibinfo{volume}{70}.
  \bibinfo{publisher}{{PMLR}}, \bibinfo{pages}{1885--1894}.
\newblock
\showURL{%
\url{http://proceedings.mlr.press/v70/koh17a.html}}


\bibitem[\protect\citeauthoryear{Le, Hu, Song, Zhang, Zhao, and Yan}{Le
  et~al\mbox{.}}{2019}]%
        {DBLP:conf/cikm/LeHSZ0019}
\bibfield{author}{\bibinfo{person}{Ran Le}, \bibinfo{person}{Wenpeng Hu},
  \bibinfo{person}{Yang Song}, \bibinfo{person}{Tao Zhang},
  \bibinfo{person}{Dongyan Zhao}, {and} \bibinfo{person}{Rui Yan}.}
  \bibinfo{year}{2019}\natexlab{}.
\newblock \showarticletitle{Towards Effective and Interpretable Person-Job
  Fitting}. In \bibinfo{booktitle}{{\em {CIKM}}},
  \bibfield{editor}{\bibinfo{person}{Wenwu Zhu}, \bibinfo{person}{Dacheng Tao},
  \bibinfo{person}{Xueqi Cheng}, \bibinfo{person}{Peng Cui},
  \bibinfo{person}{Elke~A. Rundensteiner}, \bibinfo{person}{David Carmel},
  \bibinfo{person}{Qi~He}, {and} \bibinfo{person}{Jeffrey~Xu Yu}} (Eds.).
  \bibinfo{publisher}{{ACM}}, \bibinfo{pages}{1883--1892}.
\newblock
\showDOI{%
\url{https://doi.org/10.1145/3357384.3357949}}


\bibitem[\protect\citeauthoryear{LinkedIn}{LinkedIn}{}]%
        {linkedinreport}
\bibfield{author}{\bibinfo{person}{LinkedIn}.}
\newblock \bibinfo{booktitle}{{\em LinkedIn Workforce Report|United States}}.
\newblock
\showURL{%
\url{https://economicgraph.linkedin.com/resources/linkedin-workforce-report-march-2020}}


\bibitem[\protect\citeauthoryear{Lipton}{Lipton}{2016}]%
        {DBLP:journals/corr/Lipton16a}
\bibfield{author}{\bibinfo{person}{Zachary~Chase Lipton}.}
  \bibinfo{year}{2016}\natexlab{}.
\newblock \showarticletitle{The Mythos of Model Interpretability}.
\newblock \bibinfo{journal}{{\em CoRR\/}}  \bibinfo{volume}{abs/1606.03490}
  (\bibinfo{year}{2016}).
\newblock
\showeprint[arxiv]{1606.03490}
\showURL{%
\url{http://arxiv.org/abs/1606.03490}}


\bibitem[\protect\citeauthoryear{Lu, Helou, and Gillet}{Lu
  et~al\mbox{.}}{2013}]%
        {DBLP:conf/www/LuHG13}
\bibfield{author}{\bibinfo{person}{Yao Lu}, \bibinfo{person}{Sandy~El Helou},
  {and} \bibinfo{person}{Denis Gillet}.} \bibinfo{year}{2013}\natexlab{}.
\newblock \showarticletitle{A recommender system for job seeking and recruiting
  website}. In \bibinfo{booktitle}{{\em {WWW}}}.
  \bibinfo{publisher}{International World Wide Web Conferences Steering
  Committee / {ACM}}, \bibinfo{pages}{963--966}.
\newblock
\showDOI{%
\url{https://doi.org/10.1145/2487788.2488092}}


\bibitem[\protect\citeauthoryear{{Malinowski}, {Keim}, {Wendt}, and
  {Weitzel}}{{Malinowski} et~al\mbox{.}}{2006}]%
        {1579569}
\bibfield{author}{\bibinfo{person}{J. {Malinowski}}, \bibinfo{person}{T.
  {Keim}}, \bibinfo{person}{O. {Wendt}}, {and} \bibinfo{person}{T. {Weitzel}}.}
  \bibinfo{year}{2006}\natexlab{}.
\newblock \showarticletitle{Matching People and Jobs: A Bilateral
  Recommendation Approach}. In \bibinfo{booktitle}{{\em Proceedings of the 39th
  Annual Hawaii International Conference on System Sciences (HICSS'06)}},
  Vol.~\bibinfo{volume}{6}. \bibinfo{pages}{137c--137c}.
\newblock


\bibitem[\protect\citeauthoryear{Mitra, Diaz, and Craswell}{Mitra
  et~al\mbox{.}}{2017}]%
        {DBLP:conf/www/Mitra0C17}
\bibfield{author}{\bibinfo{person}{Bhaskar Mitra}, \bibinfo{person}{Fernando
  Diaz}, {and} \bibinfo{person}{Nick Craswell}.}
  \bibinfo{year}{2017}\natexlab{}.
\newblock \showarticletitle{Learning to Match using Local and Distributed
  Representations of Text for Web Search}. In \bibinfo{booktitle}{{\em {WWW}}},
  \bibfield{editor}{\bibinfo{person}{Rick Barrett}, \bibinfo{person}{Rick
  Cummings}, \bibinfo{person}{Eugene Agichtein}, {and} \bibinfo{person}{Evgeniy
  Gabrilovich}} (Eds.). \bibinfo{publisher}{{ACM}},
  \bibinfo{pages}{1291--1299}.
\newblock
\showDOI{%
\url{https://doi.org/10.1145/3038912.3052579}}


\bibitem[\protect\citeauthoryear{Qin, Zhu, Xu, Zhu, Jiang, Chen, and Xiong}{Qin
  et~al\mbox{.}}{2018}]%
        {DBLP:conf/sigir/QinZXZJCX18}
\bibfield{author}{\bibinfo{person}{Chuan Qin}, \bibinfo{person}{Hengshu Zhu},
  \bibinfo{person}{Tong Xu}, \bibinfo{person}{Chen Zhu}, \bibinfo{person}{Liang
  Jiang}, \bibinfo{person}{Enhong Chen}, {and} \bibinfo{person}{Hui Xiong}.}
  \bibinfo{year}{2018}\natexlab{}.
\newblock \showarticletitle{Enhancing Person-Job Fit for Talent Recruitment: An
  Ability-aware Neural Network Approach}. In \bibinfo{booktitle}{{\em
  {SIGIR}}}. \bibinfo{publisher}{{ACM}}, \bibinfo{pages}{25--34}.
\newblock
\showDOI{%
\url{https://doi.org/10.1145/3209978.3210025}}


\bibitem[\protect\citeauthoryear{Ramanath, Inan, Polatkan, Hu, Guo, Ozcaglar,
  Wu, Kenthapadi, and Geyik}{Ramanath et~al\mbox{.}}{2018}]%
        {DBLP:conf/cikm/RamanathIPHGOWK18}
\bibfield{author}{\bibinfo{person}{Rohan Ramanath}, \bibinfo{person}{Hakan
  Inan}, \bibinfo{person}{Gungor Polatkan}, \bibinfo{person}{Bo Hu},
  \bibinfo{person}{Qi Guo}, \bibinfo{person}{Cagri Ozcaglar},
  \bibinfo{person}{Xianren Wu}, \bibinfo{person}{Krishnaram Kenthapadi}, {and}
  \bibinfo{person}{Sahin~Cem Geyik}.} \bibinfo{year}{2018}\natexlab{}.
\newblock \showarticletitle{Towards Deep and Representation Learning for Talent
  Search at LinkedIn}. In \bibinfo{booktitle}{{\em {CIKM}}}.
  \bibinfo{publisher}{{ACM}}, \bibinfo{pages}{2253--2261}.
\newblock
\showDOI{%
\url{https://doi.org/10.1145/3269206.3272030}}


\bibitem[\protect\citeauthoryear{Selvaraju, Cogswell, Das, Vedantam, Parikh,
  and Batra}{Selvaraju et~al\mbox{.}}{2017}]%
        {DBLP:conf/iccv/SelvarajuCDVPB17}
\bibfield{author}{\bibinfo{person}{Ramprasaath~R. Selvaraju},
  \bibinfo{person}{Michael Cogswell}, \bibinfo{person}{Abhishek Das},
  \bibinfo{person}{Ramakrishna Vedantam}, \bibinfo{person}{Devi Parikh}, {and}
  \bibinfo{person}{Dhruv Batra}.} \bibinfo{year}{2017}\natexlab{}.
\newblock \showarticletitle{Grad-CAM: Visual Explanations from Deep Networks
  via Gradient-Based Localization}. In \bibinfo{booktitle}{{\em {ICCV}}}.
  \bibinfo{publisher}{{IEEE} Computer Society}, \bibinfo{pages}{618--626}.
\newblock
\showDOI{%
\url{https://doi.org/10.1109/ICCV.2017.74}}


\bibitem[\protect\citeauthoryear{Shan, Hoens, Jiao, Wang, Yu, and Mao}{Shan
  et~al\mbox{.}}{2016}]%
        {DBLP:conf/kdd/ShanHJWYM16}
\bibfield{author}{\bibinfo{person}{Ying Shan}, \bibinfo{person}{T.~Ryan Hoens},
  \bibinfo{person}{Jian Jiao}, \bibinfo{person}{Haijing Wang},
  \bibinfo{person}{Dong Yu}, {and} \bibinfo{person}{J.~C. Mao}.}
  \bibinfo{year}{2016}\natexlab{}.
\newblock \showarticletitle{Deep Crossing: Web-Scale Modeling without Manually
  Crafted Combinatorial Features}. In \bibinfo{booktitle}{{\em {SIGKDD}}}.
  \bibinfo{publisher}{{ACM}}, \bibinfo{pages}{255--262}.
\newblock
\showDOI{%
\url{https://doi.org/10.1145/2939672.2939704}}


\bibitem[\protect\citeauthoryear{Sutskever, Vinyals, and Le}{Sutskever
  et~al\mbox{.}}{2014}]%
        {DBLP:conf/nips/SutskeverVL14}
\bibfield{author}{\bibinfo{person}{Ilya Sutskever}, \bibinfo{person}{Oriol
  Vinyals}, {and} \bibinfo{person}{Quoc~V. Le}.}
  \bibinfo{year}{2014}\natexlab{}.
\newblock \showarticletitle{Sequence to Sequence Learning with Neural
  Networks}. In \bibinfo{booktitle}{{\em {NIPS}}}. \bibinfo{pages}{3104--3112}.
\newblock


\bibitem[\protect\citeauthoryear{Tang, Lu, Liu, Mou, Vechtomova, and Lin}{Tang
  et~al\mbox{.}}{2019}]%
        {DBLP:journals/corr/abs-1903-12136}
\bibfield{author}{\bibinfo{person}{Raphael Tang}, \bibinfo{person}{Yao Lu},
  \bibinfo{person}{Linqing Liu}, \bibinfo{person}{Lili Mou},
  \bibinfo{person}{Olga Vechtomova}, {and} \bibinfo{person}{Jimmy Lin}.}
  \bibinfo{year}{2019}\natexlab{}.
\newblock \showarticletitle{Distilling Task-Specific Knowledge from {BERT} into
  Simple Neural Networks}.
\newblock \bibinfo{journal}{{\em CoRR\/}}  \bibinfo{volume}{abs/1903.12136}
  (\bibinfo{year}{2019}).
\newblock
\showeprint[arxiv]{1903.12136}


\bibitem[\protect\citeauthoryear{Vaswani, Shazeer, Parmar, Uszkoreit, Jones,
  Gomez, Kaiser, and Polosukhin}{Vaswani et~al\mbox{.}}{2017}]%
        {DBLP:conf/nips/VaswaniSPUJGKP17}
\bibfield{author}{\bibinfo{person}{Ashish Vaswani}, \bibinfo{person}{Noam
  Shazeer}, \bibinfo{person}{Niki Parmar}, \bibinfo{person}{Jakob Uszkoreit},
  \bibinfo{person}{Llion Jones}, \bibinfo{person}{Aidan~N. Gomez},
  \bibinfo{person}{Lukasz Kaiser}, {and} \bibinfo{person}{Illia Polosukhin}.}
  \bibinfo{year}{2017}\natexlab{}.
\newblock \showarticletitle{Attention is All you Need}. In
  \bibinfo{booktitle}{{\em {NIPS}}}. \bibinfo{pages}{5998--6008}.
\newblock


\bibitem[\protect\citeauthoryear{Wang, Zhang, Posse, and Bhasin}{Wang
  et~al\mbox{.}}{2013}]%
        {DBLP:conf/www/WangZPB13}
\bibfield{author}{\bibinfo{person}{Jian Wang}, \bibinfo{person}{Yi Zhang},
  \bibinfo{person}{Christian Posse}, {and} \bibinfo{person}{Anmol Bhasin}.}
  \bibinfo{year}{2013}\natexlab{}.
\newblock \showarticletitle{Is it time for a career switch?}. In
  \bibinfo{booktitle}{{\em {WWW}}}. \bibinfo{publisher}{International World
  Wide Web Conferences Steering Committee / {ACM}},
  \bibinfo{pages}{1377--1388}.
\newblock
\showDOI{%
\url{https://doi.org/10.1145/2488388.2488509}}


\bibitem[\protect\citeauthoryear{Xu, Yu, Yang, Xiong, and Zhu}{Xu
  et~al\mbox{.}}{2019}]%
        {DBLP:journals/tkde/XuYYXZ19}
\bibfield{author}{\bibinfo{person}{Huang Xu}, \bibinfo{person}{Zhiwen Yu},
  \bibinfo{person}{Jingyuan Yang}, \bibinfo{person}{Hui Xiong}, {and}
  \bibinfo{person}{Hengshu Zhu}.} \bibinfo{year}{2019}\natexlab{}.
\newblock \showarticletitle{Dynamic Talent Flow Analysis with Deep Sequence
  Prediction Modeling}.
\newblock \bibinfo{journal}{{\em {IEEE} Trans. Knowl. Data Eng.\/}}
  \bibinfo{volume}{31}, \bibinfo{number}{10} (\bibinfo{year}{2019}),
  \bibinfo{pages}{1926--1939}.
\newblock
\showDOI{%
\url{https://doi.org/10.1109/TKDE.2018.2873341}}


\bibitem[\protect\citeauthoryear{Yan, Le, Song, Zhang, Zhang, and Zhao}{Yan
  et~al\mbox{.}}{2019}]%
        {DBLP:conf/kdd/YanLSZZ019}
\bibfield{author}{\bibinfo{person}{Rui Yan}, \bibinfo{person}{Ran Le},
  \bibinfo{person}{Yang Song}, \bibinfo{person}{Tao Zhang},
  \bibinfo{person}{Xiangliang Zhang}, {and} \bibinfo{person}{Dongyan Zhao}.}
  \bibinfo{year}{2019}\natexlab{}.
\newblock \showarticletitle{Interview Choice Reveals Your Preference on the
  Market: To Improve Job-Resume Matching through Profiling Memories}. In
  \bibinfo{booktitle}{{\em {SIGKDD}}}. \bibinfo{publisher}{{ACM}},
  \bibinfo{pages}{914--922}.
\newblock
\showDOI{%
\url{https://doi.org/10.1145/3292500.3330963}}


\bibitem[\protect\citeauthoryear{Zhang, Yang, and Niu}{Zhang
  et~al\mbox{.}}{2015}]%
        {iscid2014}
\bibfield{author}{\bibinfo{person}{Yingya Zhang}, \bibinfo{person}{Cheng Yang},
  {and} \bibinfo{person}{Zhixiang Niu}.} \bibinfo{year}{2015}\natexlab{}.
\newblock \showarticletitle{A Research of Job Recommendation System Based on
  Collaborative Filtering}.
\newblock \bibinfo{journal}{{\em {ISCID}\/}}  \bibinfo{volume}{1}
  (\bibinfo{date}{03} \bibinfo{year}{2015}), \bibinfo{pages}{533--538}.
\newblock


\bibitem[\protect\citeauthoryear{Zhu, Zhu, Xiong, Ma, Xie, Ding, and Li}{Zhu
  et~al\mbox{.}}{2018}]%
        {acmtmis}
\bibfield{author}{\bibinfo{person}{Chen Zhu}, \bibinfo{person}{Hengshu Zhu},
  \bibinfo{person}{Hui Xiong}, \bibinfo{person}{Chao Ma}, \bibinfo{person}{Fang
  Xie}, \bibinfo{person}{Pengliang Ding}, {and} \bibinfo{person}{Pan Li}.}
  \bibinfo{year}{2018}\natexlab{}.
\newblock \showarticletitle{Person-Job Fit: Adapting the Right Talent for the
  Right Job with Joint Representation Learning}.
\newblock \bibinfo{journal}{{\em ACM Transactions on Management Information
  Systems\/}}  \bibinfo{volume}{9} (\bibinfo{date}{09} \bibinfo{year}{2018}),
  \bibinfo{pages}{1--17}.
\newblock
\showDOI{%
\url{https://doi.org/10.1145/3234465}}


\end{thebibliography}

\end{document}